\documentclass[preprint]{revtex4}
\usepackage{amsmath}
\usepackage{mathrsfs}

\usepackage{amssymb}
\usepackage{graphicx}
\usepackage{subfigure}
\usepackage[colorlinks,
            linkcolor=blue,
            anchorcolor=blue,
            citecolor=blue]{hyperref}
\newtheorem{theorem}{Theorem}

\def\ra{\rangle}

\allowdisplaybreaks[4]

\begin{document}

\title{Product and sum uncertainty relations based on\\ metric-adjusted skew information}

\author{Xiaoyu Ma$^{1}$}
\author{Qing-Hua Zhang$^{2,}$\footnotemark[1]}
\author{Shao-Ming Fei$^{2,3,}$\footnotemark[1]}

\affiliation{$^1$Zhongtai Securities Institute for Financial Studies, Shandong University, Jinan 250100, China \\
$^2$School of Mathematical Sciences, Capital Normal University,
Beijing 100048, China\\
$^3$Max-Planck-Institute for Mathematics in the Sciences, 04103 Leipzig, Germany}

\renewcommand{\thefootnote}{\fnsymbol{footnote}}


\bigskip
\begin{abstract}
The metric-adjusted skew information establishes a connection between the geometrical formulation of quantum statistics and the measures of quantum information. We study uncertainty relations in product and summation forms of metric-adjusted skew information. We present lower bounds on product and summation uncertainty inequalities based on metric-adjusted skew information via operator representation of observables. Explicit examples are provided to back our claims.
\end{abstract}

\maketitle

\section{Introduction}\label{sec1}

In the theory of quantum physics, there are limitations on the measurement of quantum mechanical observables which are not commutative with certain conserved quantities due to the conservation law \cite{Wigner1952Die,araki1960measurement}. Unlike the uncertainty formulation based on quantum variance proposed by Heisenberg and Robertson \cite{Heisenberg1927,PhysRev.34.163}, Luo shew that the skew information provides us with a new notion to quantify Bohr's complementarity principle and the Heisenberg uncertainty principle \cite{Luo2003Wigner}. Here the Wigner-Yanase skew information of a quantum state $\rho$ with respect to an observable $A$ is given by \cite{Wigner1963INFORMATION},
\begin{equation}
I_{\rho}(A)=-\frac{1}{2}{\rm Tr}([\sqrt{\rho},A]^2).
\end{equation}
$I_{\rho}(A)$ can be interpreted as a measure of non-commutativity between square root of the state $\rho$ and the conserved observable $A$. Wigner and Yanase proved that this quantity satisfies all the desirable requirements of an information measure.
In Ref.~\cite{Luo2003Wigner}, it has been shown that such Wigner-Yanase skew information satisfies the following uncertainty inequality,
\begin{equation}
   I_{\rho}(A)I_{\rho}(B)\geq \frac{1}{4}|{\rm Tr}([\sqrt{\rho},A][\sqrt{\rho},B])|^2\geq\frac{1}{4}|{\rm Tr}(\rho [A,B])|^2.
\end{equation}

Later, Dyson  generalized the skew information to the called Wigner-Yanase-Dyson skew information which is shown to be convex by Lieb \cite{LIEB1973267}. On the other hand, from the quantum counterpart of Cramr\'er-Rao inequality \cite{holevo2011probabilistic,Braunstein1994Statistical} the quantum Fisher information was also introduced based on symmetric logarithmic derivative \cite{helstrom1969quantum}. In fact, these kinds of information are the particular cases of the metric-adjusted skew information. Let $M_d(\mathbb{C})$ be the complex matrix space and $S_d(\mathbb{C})$ the $d$-dimensional density matrices. For $A,B\in M_d(\mathbb{C})$ and $\rho\in S_d(\mathbb{C})$, the monotone metric $K^c_\rho(\cdot,\cdot)$ is defined by 
$K^c_\rho(A,B)={\rm Tr}(A^\dagger c(L_\rho,R_\rho)B)$, where $c(L_\rho,R_\rho)$ is called Morozova-Chentsov function with respect to the left and right multiplication operators $L_\rho$ and $R_\rho$ \cite{PETZ199681}. The metric $K^c_\rho(\cdot,\cdot)$ satisfies the following conditions:
(a) $K^c_\rho(\cdot,\cdot)$ is sesquilinear; (b) $K^c_\rho(A,A)\geq 0$ and the equality holds if and only if $A=0$; (c) $K^c_\rho(\cdot,\cdot)$ is continuous on $S_d(\mathbb{C})$ for every $\rho$; (d) $K^c_{T(\rho)}({\rm T}(A),{\rm T}(A))\leq K^c_\rho(A,A)$ for every stochastic map ${\rm T}: M_d(\mathbb{C})\rightarrow M_D(\mathbb{C})$.
While the Morozova-Chentsov function is given by a positive operator monotone function $f$,
\begin{equation}
c(x,y)=\frac{1}{yf(xy^{-1})},\qquad x,\ y>0,
\end{equation}
where $f$ satisfies the functional equation, $f(t)=tf(t^{-1})$, $t>0$.

The metric-adjusted skew information $I_{\rho}^{c}(A)$ for any quantum state $\rho $ with respect to an observable $A$ is defined by the symmetric monotone metric \cite{Hansen9909,GIBILISCO20092225},
\begin{equation}
\begin{aligned}
I_{\rho}^{c}(A) &=\frac{m(c)}{2} K_{\rho}^{c}(i[\rho, A], i[\rho, A]) \\
&=\frac{m(c)}{2} \operatorname{Tr} (i[\rho, A] c\left(L_{\rho}, R_{\rho}\right) i[\rho, A]),
\end{aligned}
\end{equation}
where $m(c)=\lim_{t\rightarrow 0}f(t)$. The metric-adjusted skew
information has innumerable different formulations corresponding to different Morozova-Chentsov functions. The quantum Fisher information, the Wigner-Yanase skew information and the Wigner-Yanase-Dyson skew information are special cases of the metric-adjusted skew information $I_{\rho}^{c}(A)$.
For instance, if one takes the function $f$ to be
\begin{equation}
f_\alpha=\alpha(1-\alpha)\frac{(1-t)^2}{(1-t^\alpha)(1-t^{1-\alpha})},\qquad t>0,
\end{equation}
with $f(0)=\alpha(1-\alpha)$, the corresponding Morozova-Chentsov function $c$ becomes
\begin{equation}\label{dysonmetric}
c_{\alpha}(x, y)=\frac{1}{\alpha(1-\alpha)} \frac{\left(x^{\alpha}-y^{\alpha}\right)\left(x^{1-\alpha}-y^{1-\alpha}\right)}{(x-y)^{2}},\qquad 0<\alpha<1,
\end{equation}
and the metric-adjusted skew information $I_{\rho}^{{\alpha}}(A)$ becomes the Wigner-Yanase-Dyson skew information,
\begin{equation}\label{WYDSI}
\begin{aligned}
I_{\rho}^{c_{\alpha}}(A) &=\frac{\alpha(1-\alpha)}{2} \operatorname{Tr}\left\{i[\rho, A] c_{\alpha}\left(L_{\rho}, R_{\rho}\right) i[\rho, A]\right\} \\
&=-\frac{1}{2} \operatorname{Tr}\left[\rho^{\alpha}, A\right]\left[\rho^{1-\alpha}, A\right].
\end{aligned}
\end{equation}
If $\alpha=1/2$, $I_{\rho}^{c_{\alpha}}(A)$ becomes the Wigner-Yanase skew information $I_{\rho}(A)$.

Recently, Cai generalized sum uncertainty relations of Wigner-Yanase skew information to metric-adjusted skew information, and established a series of lower bounds given by the skew information of any prescribed size of the combinations \cite{Cai2021Sum}. Subsequently, Ren $et\ al.$ supplemented the sum uncertainty relations based on metric-adjusted skew information by using the properties of matrix norm \cite{PhysRevA.104.052414}. This paper focuses on quantum uncertainty relations in the forms of product and summation of metric-adjusted skew information. We establish the product uncertainty relations of the metric-adjusted skew information with the help of two different refinements of the Cauchy-Schwarz inequality in Sec.~\ref{sec3}. We propose the sum uncertainty relations based on the operator representation of observables in Sec.~\ref{sec4}. We summarize our results in Sec.~\ref{sec5}.

\section{Uncertainty Relations in Product Form}\label{sec3}

In this section, we study stronger sum uncertainty relations based on the operator representation of the observables. We consider $d$-dimension quantum system whose Hilbert space is spanned by computation basis $|i\ra,\ i=1,2,\dots,d$. The complete set of local orthogonal observables (LOOs) is a set of $d^2$ observables $\Omega_\mu$, satisfying orthogonal relations ${\rm Tr}(\Omega_\mu \Omega_\nu)=\delta_{\mu\nu}$, $\mu,\nu=1,2,\dots,d^2$. They form an orthogonal basis of all observables, that is, any observable $A$ has an expansion,
\begin{equation}
A=\sum_\mu {\rm Tr}(\Omega_\mu A)\Omega_\mu=\sum_\mu a_\mu \Omega_\mu\equiv\vec{a}^T\vec{\Omega},
\end{equation}
where $\vec{a}^T$ denotes the transpose of the vector $\vec{a}=(a_1,a_2,...,a_{d^2})$. We denote the correlation measure of observables $A$ and $B$ by
\begin{equation}
    Corr_{\rho}^c(A,B)=\frac{m(c)}{2} K_{\rho}^{c}(i[\rho, A], i[\rho, B]).
\end{equation}

Consider two quantum observables $A=\vec{a}^T\vec{\Omega}$ and $B=\vec{b}^T\vec{\Omega}$. The metric-adjusted skew information of $A$ is given by
\begin{equation}
I^{c}_{\rho}(A)=Corr_{\rho}^c(A,A)=\frac{m(c)}{2} K_{\rho}^{c}(i[\rho, A], i[\rho, A])\equiv\vec{a}^T\Gamma\vec{a},
\end{equation}
where $\Gamma$ is a positive semi-definite matrix with entries $\Gamma_{\mu\nu}=Corr_{\rho}^c(\Omega_\mu,\Omega_\nu)$. For given $\Gamma$ there exists a matrix $C$ such that $\Gamma=C^\dagger C$. Hence, the skew information can be rewritten as:
\begin{equation}
I^{c}_{\rho}(A)=\vec{a}^TC^\dagger C\vec{a}\equiv |\vec{f}|^2,
\end{equation}
where $\vec{f}=(\alpha_1.\alpha_2,\cdots,\alpha_{d^2})^T=C\vec{a}$. Similarly, $I^{c}_{\rho}(B)\equiv |\vec{g}|^2$ with $\vec{g}=(\beta_1.\beta_2,\cdots,\beta_{d^2})^T=C\vec{b}$. The product of skew information of these two observables is given by \cite{PhysRevA.100.022116}
\begin{equation}\label{cauchy}
\begin{aligned}
      I^{c}_{\rho}(A)I^{c}_{\rho}(B)
      =&|\vec{f}|^2|\vec{g}|^2=\sum_{i}|\alpha_i|^2\sum_{j}|\beta_j|^2 \\
      \geq &|\sum_{i}\alpha_i^*\beta_i|^2=|\vec{f}^\dagger \vec{g}|^2=|Corr_{\rho}^c(A,B)|^2,
\end{aligned}
\end{equation}
where the first inequality is due to the Cauchy-Schwarz inequality.

Let $\vec{X}=(x_1,x_2,\dots,x_{d^2})^T$ and $x_i=|\alpha_i|$ be non negative real number. Similarly, let $\vec{Y}=(y_1,y_2,\dots,y_{d^2})^T$ with $y_i=|\beta_i|$.
Define the refinement of Cauchy-Schwarz inequality by the geometric-arithmetic mean inequality:
\begin{equation}\label{refine1}
\begin{aligned}
I_{k}= \sum_{1 \leq i \leq d^2} x_{i}^{2} y_{i}^{2}+\sum_{\genfrac{}{}{0pt}{}{k<j}{1 \leq i<j \leq d^2}}\left(x_{i}^{2} y_{j}^{2}+x_{j}^{2} y_{i}^{2}\right) +\sum_{1 \leq i<j \leq k} 2 x_{i} y_{i} x_{j} y_{j}.
\end{aligned}
\end{equation}
Note that $I_1=I^{c}_{\rho}(A)I^{c}_{\rho}(B)$ and $I_{d^2}= (\sum_{i}x_iy_i)^2$. The refinement is a descending sequence,
\begin{equation}
I_{k+1}-I_{k}=-\left(\sum_{i=1}^{k} x_{i} y_{k+1}+y_{i} x_{k+1}\right)^{2} \leqslant 0,
\end{equation}
namely, $I_1\geq I_2\geq \dots\geq I_{d^2}$.
Naturally, we have the following product uncertainty relations via metric-adjusted skew information.

\begin{theorem}\label{th1}
Let A and B be two arbitrary observables on $d$-dimension Hilbert space. The product of the metric-adjusted skew information of A and B satisfies the following uncertainty relations:
\begin{equation}
I^{c}_{\rho}(A)I^{c}_{\rho}(B) \geq I_k,
\end{equation}
where $k=1,2,\dots,d^2$.
\end{theorem}

Moreover, since for arbitrary two $d^2$-element permutations $\pi_A,\pi_B\in S(d^2)$, it always holds that
\begin{equation}
I^{c}_{\rho}(A)I^{c}_{\rho}(B)=\sum_{i}x_{\pi_A(i)}^2\sum_{j}y_{\pi_B(j)}^2,
\end{equation}
we can define other refinements corresponding to different pairs of permutations $(\pi_A,\pi_B)$,
\begin{equation}
\begin{aligned}
(\pi_A,\pi_B)I_k=
&\sum_{1 \leq i \leq d^2} x_{{\pi_A(i)}}^{2} y_{\pi_B(i)}^{2}+\sum_{\genfrac{}{}{0pt}{}{k<j}{1 \leq i<j \leq d^2}}\left(x_{\pi_A(i)}^{2} y_{\pi_B(j)}^{2}+x_{\pi_A(j)}^{2} y_{\pi_B(i)}^{2}\right) \\
& +\sum_{1 \leq i<j \leq k} 2 x_{\pi_A(i)} y_{\pi_B(i)} x_{\pi_A(j)} y_{\pi_B(j)}.
\end{aligned}
\end{equation}
Then we have the following general uncertainty relations of metric-adjusted skew information under element permutations,
\begin{equation}
I^{c}_{\rho}(A)I^{c}_{\rho}(B) \geq \max_{\pi_A,\pi_B\in S(d^2)}(\pi_A,\pi_B)I_k.
\end{equation}

Based on the sequence (\ref{refine1}), the authors in Ref.~\cite{Li2019An} provided a more refined descending sequence in studying uncertainty relations related to unitary operators based on deviations. For each $p \geq 3$ and $q=1,2, \cdots,(p-1)$, define
\begin{equation}\label{refine2}
S_{p q}=-\sum_{j=2}^{p-1} \sum_{i=1}^{j-1}\left(x_{j} y_{i}-x_{i} y_{j}\right)^{2}-\sum_{m=1}^{q}\left(x_{p} y_{m}-x_{m} y_{p}\right)^{2}+\sum_{i, j}^{d^2} x_{i}^{2} y_{j}^{2}.
\end{equation}
In particular, $S_{10}=\sum_{i, j}^{d^2} x_{i}^{2} y_{j}^{2}$ and $S_{21}=\sum_{1 \leq i \leq d^2} x_{i}^{2} y_{i}^{2}+\sum_{1 \leq i<j \leq d^2,2<j}\left(x_{i}^{2} y_{j}^{2}+x_{j}^{2} y_{i}^{2}\right)+2 x_{1} y_{1} x_{2} y_{2}$. The sequence is a descending refinement satisfying
$$
\begin{aligned}
&S_{21}-S_{10}=-\left(x_{2} y_{1}-x_{1} y_{2}\right)^{2} \leq 0, \\
&S_{p q}-S_{p(q-1)}=-\left(x_{p} y_{q}-y_{q} x_{p}\right)^{2} \leq 0, \\
&S_{p 1}-S_{(p-1)(p-2)}=-\left(x_{p} y_{1}-x_{1} y_{p}\right)^{2} \leq 0.
\end{aligned}
$$
That is to say,
\begin{equation}
   S_{10} \geq S_{21} \geq S_{31} \geq S_{32} \geq S_{41} \geq \cdots \geq S_{d^2 1} \geq S_{d^2 2} \geq S_{d^2 3} \geq \cdots \geq S_{d^2(d^2-1)}.
\end{equation}
The descending sequence (\ref{refine2}) is a refinement of the sequence (\ref{refine1}). When one takes $q=p-1$,
\begin{equation}
   \begin{aligned}
S_{p(p-1)} &=-\sum_{j=2}^{p-1} \sum_{i=1}^{j-1}\left(x_{j} y_{i}-x_{i} y_{j}\right)^{2}-\sum_{m=1}^{p-1}\left(x_{p} y_{m}-x_{m} y_{p}\right)^{2}+\sum_{i, j}^{d^2} x_{i}^{2} y_{j}^{2} \\
&=\sum_{1 \leq i \leq d^2} x_{i}^{2} y_{i}^{2}+\sum_{\genfrac{}{}{0pt}{}{1 \leq i<j\leq d^2}{p<j}}\left(x_{i}^{2} y_{j}^{2}+x_{j}^{2} y_{i}^{2}\right)+\sum_{1 \leq i<j \leq p} 2 x_{i} y_{i} x_{j} y_{j}=I_{p}.
\end{aligned}
\end{equation}
Based on the sequence (\ref{refine2}), we have

\begin{theorem}\label{th2}
The following uncertainty relations hold for two arbitrary observables  A and B,
\begin{equation}\label{thm2}
I^{c}_{\rho}(A)I^{c}_{\rho}(B) \geq S_{pq}.
\end{equation}
\end{theorem}

Similarly, Theorem 2 has a more general form if we consider the refinements of $S_{pq}$ corresponding to different pairs of permutations $(\pi_A,\pi_B)$.
\begin{equation}
\begin{aligned}
(\pi_A,\pi_B)      S_{p q}=&-\sum_{j=2}^{p-1} \sum_{i=1}^{j-1}\left(x_{\pi_A(j)} y_{\pi_B(i)}-x_{\pi_A(i)} y_{\pi_B(j)}\right)^{2}\\
&-\sum_{m=1}^{q}\left(x_{\pi_A(p)} y_{\pi_B(m)}-x_{\pi_A(m)} y_{\pi_B(p)}\right)^{2}
+\sum_{i, j}^{d^2} x_{\pi_A(i)}^{2} y_{\pi_B(j)}^{2}.
\end{aligned}
\end{equation}
Thus in general we have the following inequality,
\begin{equation}
I^{c}_{\rho}(A)I^{c}_{\rho}(B) \geq \max_{\pi_A,\pi_B\in S(d^2)}(\pi_A,\pi_B)S_{pq}.
\end{equation}

\begin{figure}[htbp]
\centering
\includegraphics[width=10cm]{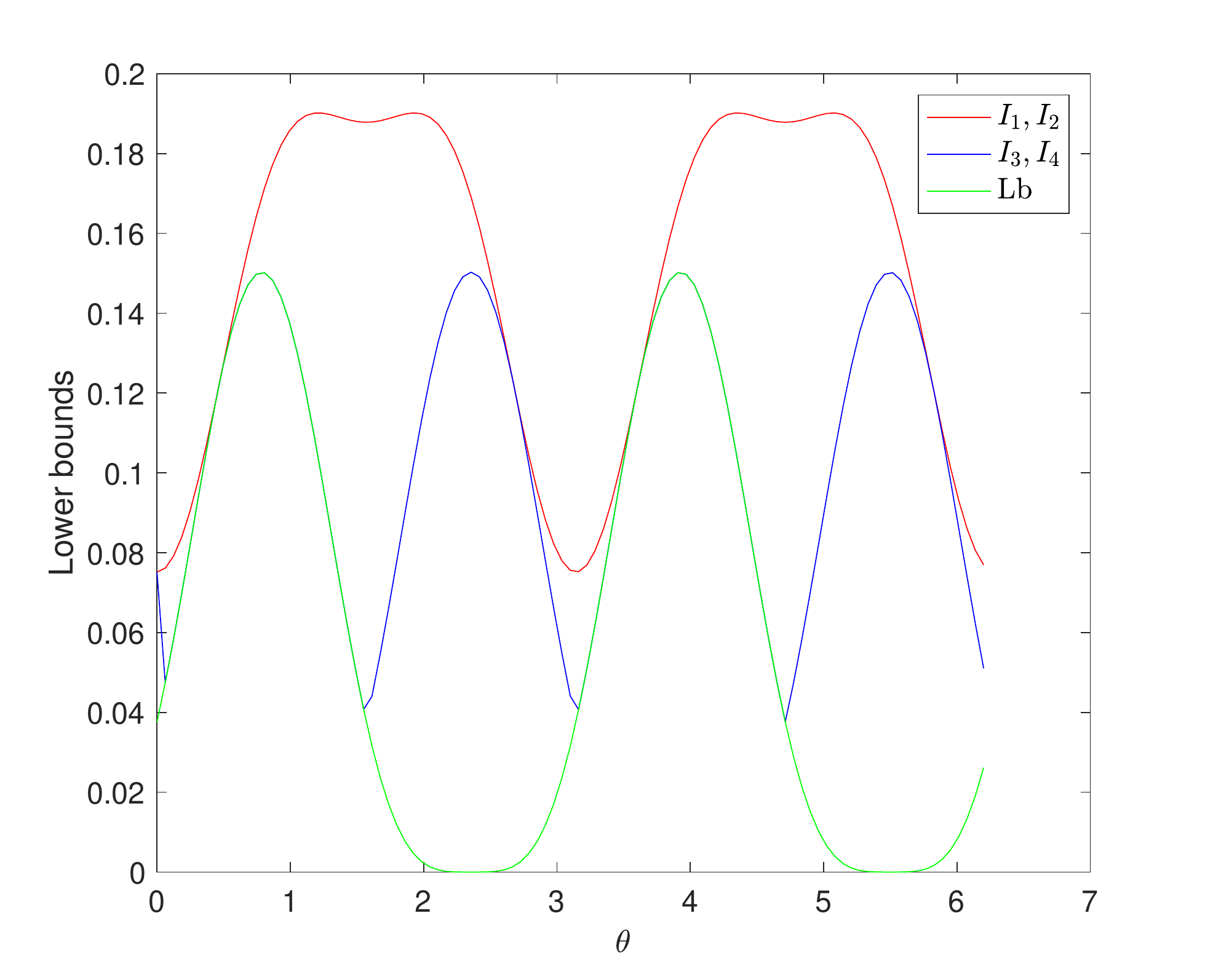}
\caption{{\rm Lb} denots the lower bound $|Corr_{\rho}^c(A,B)|^2$. The figure shows that the lower bounds of Theorem \ref{th2} cover that of Theorem \ref{th1} and both of them are tighter than that of (\ref{cauchy}).}
 \label{figex1}
\end{figure}

\emph{Example 1} Let us consider the mixed state given by the Bloch vector $\vec{r}=(\frac{\sqrt{3}}{2}cos\theta,\frac{\sqrt{3}}{2}sin\theta,0)$ \cite{Zhang2021A},
\begin{equation}                 \label{ex1}
\rho=\frac{I_2+\vec{r}\cdot\vec{\sigma}}{2},
\end{equation}
where $\vec{\sigma}=(\sigma_x,\sigma_y,\sigma_z)$ is given by the standard Pauli matrices, $I_2$ is the $2\times 2$ identity matrix. 
We take two observables $A=\sigma_x-\sigma_z/2$ and $B=\sigma_y+\sigma_z$ and use the Wigner-Yanase-Dyson skew information (\ref{WYDSI}) ($\alpha=1/4$) to illustrate the lower bounds of Theorem \ref{th1} and Theorem \ref{th2}, see Fig.~\ref{figex1}. In this case, we have $I^{c_{\frac{1}{4}}}_{\rho}(A)I^{c_{\frac{1}{4}}}_{\rho}(B)=I_1=I_2=S_{10}=S_{21}$ and     $I_3=I_4=S_{31}=S_{32}=S_{41}=S_{42}=S_{43}$.

\emph{Example 2} Consider a pure qutrit state  $|\psi\rangle=\cos{\theta}|0\rangle+\sin{\theta}|2\ra$. We take the two observables to be
\begin{equation}
A=\begin{pmatrix} 1 & 1-i& 0 \\1+i& -1 &0\\0&0&0 \end{pmatrix},~~
B=\begin{pmatrix} 0  & 0& 1-i \\ 0  &0&1\\1+i &1&0  \end{pmatrix}.
\end{equation}
Consider the Wigner-Yanase-Dyson skew information with ($\alpha=1/4$). For $\theta=\pi/4$ we have
\begin{equation}
\begin{aligned}
    &I_k=S_{10}=S_{pq}=1.875,~k=1,\cdots 6,~p=1,\cdots, 7,~q=1,\cdots, 5,\\
    &I_7=S_{76}=S_{81}=S_{82}=S_{83}=S_{84}=S_{85}=1.844,~ S_{86}=1.344,\\
    &I_8=S_{87}=S_{91}=S_{92}=S_{93}=S_{94}=S_{95}=0.625,~\\
    &~S_{96}=S_{97}=0.610,~I_9=S_{98}=|Corr_{\psi}^c(A,B)|^2=0.250.
\end{aligned}
\end{equation}
We conclude that the lower bounds of Theorem \ref{th2} cover that of Theorem \ref{th1} in this case.

\section{Uncertainty Relations in Sum Form}\label{sec4}
Consider $N$ observables $\{A_i\}_{i=1}^N$ given by $A_i=\vec{a_i}^T\vec{\Omega}$. The metric-adjusted skew information of $A_i$ with respect to $\rho$ has the following form:
$I^{c}_{\rho}(A_i)= \|\vec{X_i}\|^2$, $i=1,\cdots,N$, where $\vec{X_i}=(x_{i1},x_{i2},\dots,x_{id^2})^T$ with $x_{ij}=|(C\vec{a}_i)_j|$ and $\|\cdot \|$ stands for the norm of a vector defined by inner product. We have the following stronger sum uncertainty relations based on metric-adjusted skew information for $N$ observables.

\begin{theorem}\label{th3}
Let  $A_1, A_2, \dots, A_N$ be $N$ arbitrary observables. The following metric-adjusted skew information-based sum uncertainty relation holds for any quantum state $\rho$,
\begin{equation} \label{th3eq1}
\sum_{i=1}^N I^{c}_{\rho}(A_i)  \geq \max_{\pi_{A_i},\pi_{A_j} \in S_{d^2}} \frac{1}{2N-2}\Bigg\{ \sum_{1\leq i<j\leq N} \Lambda_{\pi_{A_i}(i)\pi_{A_j}(j)}^2+\frac{2}{N(N-1)}\Big [\sum_{1\leq i<j\leq N} \bar{\Lambda}_{\pi_{A_i}(i)\pi_{A_j}(j)}\Big ]^2 \Bigg\},
\end{equation}
where
 $${\Lambda}_{\pi_{A_i}(i)\pi_{A_j}(j)}^2=\sum_{k=1}^{d^2} (x_{i,\pi_{A_i}(k)}+x_{j,\pi_{A_j}(k)})^2,$$

 $$\bar{\Lambda}_{\pi_{A_i}(i)\pi_{A_j}(j)}^2=\sum_{k=1}^{d^2} (x_{i,\pi_{A_i}(k)}-x_{j,\pi_{A_j}(k)})^2,$$
and $\pi_{A_i},\pi_{A_j} \in S_{d^2}$ are arbitrary $d^2$-element permutations.
\end{theorem}

{\sf [Proof]} By noting the parallelogram law for all vectors $\vec{X_i}$,
\begin{equation*}
(2N-2)\sum_{i=1}^{N} \| \vec{X_i}\|^2 = \sum_{1\leq i<j \leq N} \| \vec{X_i}+\vec{X_j} \|^2 + \sum_{1\leq i<j \leq N} \| \vec{X_i}-\vec{X_j} \|^2.
\end{equation*}
and using the Cauchy-Schwarz inequality, we have
\begin{equation*}
\sum_{1\leq i<j \leq N} \| \vec{X_i} - \vec{X_j} \|^2 \geq \frac{2}{N(N-1)} (\sum_{1\leq i<j \leq N} \| \vec{X_i} - \vec{X_j} \|)^2.
\end{equation*}
Therefore,
\begin{equation}\label{th3pf1}
\sum_{i=1}^{N} \| \vec{X_i}^{\pi_i} \|^2 \geq \frac{1}{2N-2}[\frac{2}{N(N-1)}(\sum_{1\leq i<j \leq N} \| \vec{X_i}^{\pi_i} - \vec{X_j}^{\pi_j}  \|)^2 + \sum_{1\leq i<j \leq N} \| \vec{X_i} ^{\pi_i} + \vec{X_j}^{\pi_j}  \|^2],
\end{equation}
where $\vec{X_i}^{\pi_i}=(x_{{i,\pi_i(1)}},x_{{i,\pi_i(2)}},\dots,x_{{i,\pi_i(d^2)}})^T$.
This completes the proof. $\Box$

In Ref.~\cite{Zhang2021Tighter}, the authors proposed sum uncertainty relations based on Wigner-Yanase skew information in terms of the properties of matrix norm. For arbitrary finite $N$ observables $A_1, A_2, \dots, A_N$, the following  sum uncertainty relations hold:
\begin{equation}\label{zhang}
\begin{aligned}
\sum_{i=1}^N I_{\rho} (A_i)\geq \max_{x\in\{0,1\}}
&\frac{1}{2N-2}\Bigg\{\frac{2}{N(N-1)} \Big [\sum_{1\leq i<j\leq N} \sqrt{I_{\rho}(A_i+(-1)^xA_j)}\Big ]^2\\
&+\sum_{1\leq i<j\leq N} I_{\rho} (A_i+(-1)^{x+1}A_j) \Bigg\}.
\end{aligned}
\end{equation}
Below we give an example to illustrate the relations between (\ref{zhang}) and the one from Theorem \ref{th3}.

\begin{figure}[htbp]
\centering
\includegraphics[width=10cm]{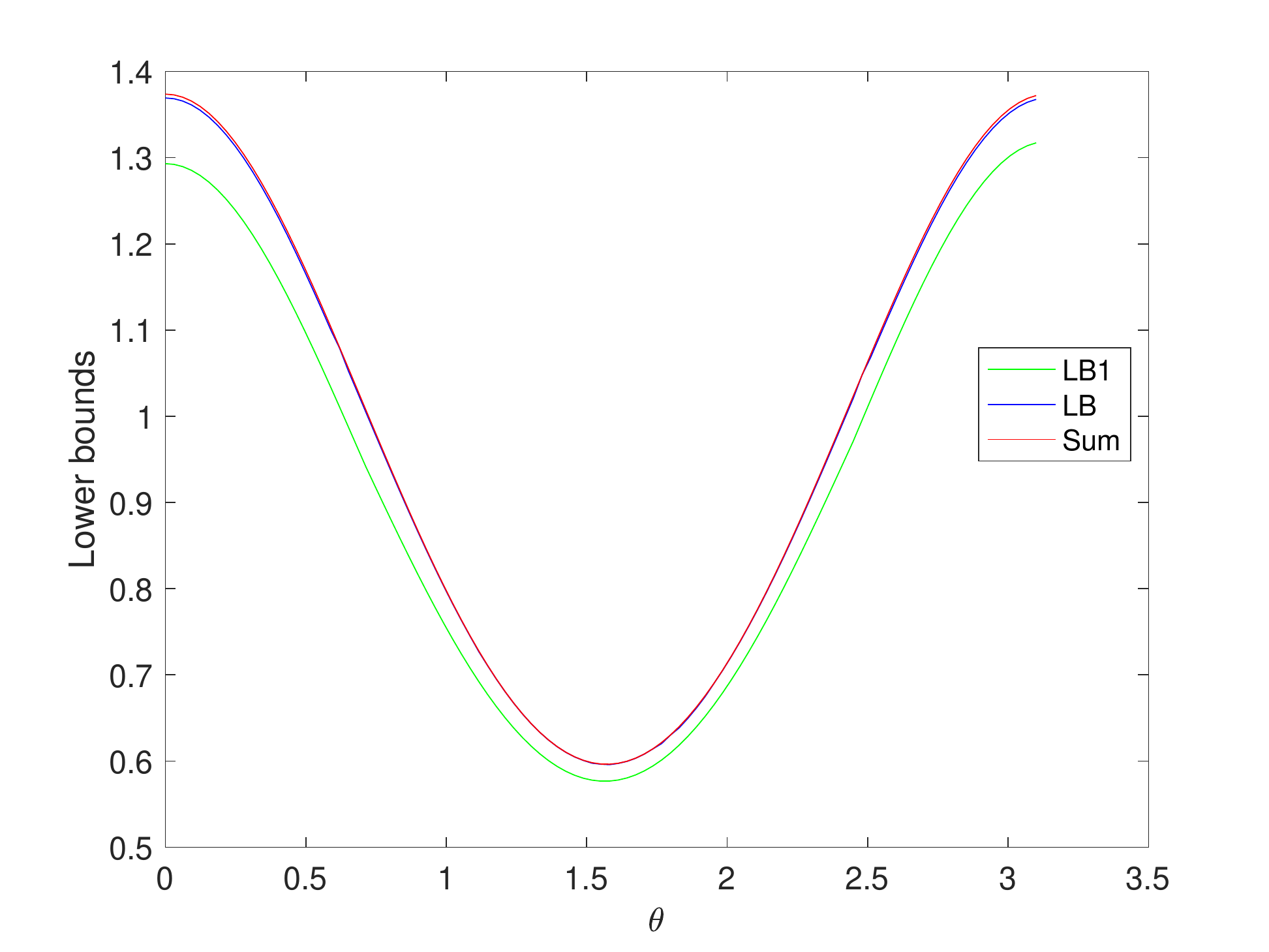}
\caption{LB1 and LB respectively represent the lower bounds of (\ref{zhang}) and (\ref{th3eq1}). Sum represents $I^{c_{\frac{1}{2}}}_\rho(A)+I^{c_{\frac{1}{2}}}_\rho(B)+I^{c_{\frac{1}{2}}}_\rho(C)$. Our lower bound LB is strictly larger than LB1.}
\label{figex3}
\end{figure}

\emph{Example 3} Consider the mixed state $\rho=\frac{1}{2}(I_2+\vec{r}\cdot\vec{\sigma})$ with $\vec{r}=(\frac{\sqrt{3}}{3}cos\theta,0,\frac{\sqrt{3}}{3})$. We use the Wigner-Yanase-Dyson skew information (\ref{WYDSI}) ($\alpha=1/2$) for comparison with the lower bound of (\ref{zhang}). We take three observables $A=\sigma_x+\sigma_y/2$, $B=\sigma_y$ and $C=\sigma_z-\sigma_y$. The figure Fig.~\ref{figex3} shows that the lower bound of (\ref{th3eq1}) is strictly larger than that of (\ref{zhang}).

\section{Conclusion}\label{sec5}
We have studied uncertainty relations based on metric-adjusted skew information of quantum observables, which include the uncertainty relations of Wigner-Yanase skew information of quantum observables as special cases. By the use of the geometric-arithmetic mean inequality, we have presented a series of uncertainty inequalities to characterize the uncertainty in product form of two observables. We have also put forward the sum uncertainty relations for $N$ quantum observables. These conclusions give rise to a new starting point for further investigations on uncertainty relations based on metric-adjusted skew information and their related implications and applications.

\bigskip
\noindent{\bf Acknowledgments}\, This work is supported by NSFC (Grant Nos. 12075159, 12171044), Beijing Natural Science Foundation (Z190005), Academy for Multidisciplinary Studies, Capital Normal University, the Academician Innovation Platform of Hainan Province, Shenzhen Institute for Quantum Science and Engineering, Southern University of Science and Technology (No. SIQSE202001), and China Scholarship Council.

\noindent{\bf Data availability}\, Data sharing not applicable to this article as no data sets were generated or analyzed during the current study.

\bibliographystyle{apsrev4-1}
\bibliography{article}
\end{document}